\documentclass[runningheads]{svmult}

\usepackage{makeidx}   
\usepackage{graphicx}  
\usepackage{subeqnar}  
\usepackage{multicol}  
\usepackage{physprbb}  
\makeindex             

\def\la{\mathrel{\hbox{\rlap{\hbox{\lower4pt\hbox{$\sim$}}}\hbox{$<$}}}}
\def\ga{\mathrel{\hbox{\rlap{\hbox{\lower4pt\hbox{$\sim$}}}\hbox{$>$}}}}

\begin{document}
\title*{Construction of the Variability $\rightarrow$ Luminosity Estimator}
\toctitle{Focusing of a Parallel Beam to Form a Point
\protect\newline in the Particle Deflection Plane}
%
%
\titlerunning{Construction of the Variability $\rightarrow$ Luminosity 
Estimator}
%
\author{D. E. Reichart\inst{1,2}
\and D. Q. Lamb\inst{3}}
\authorrunning{Reichart \& Lamb}
%
%
\institute{Department of Astronomy, California Institute of Technology,
Mail Code 105-24, 1201 East California Boulevard, Pasadena, CA 91125
\and Hubble Fellow
\and Department of Astronomy and Astrophysics, University of
Chicago, 5640 South Ellis Avenue, Chicago, IL 60637}

\maketitle              

\section{Introduction}

In this paper, we present a possible luminosity estimator for the long-duration
bursts, the construction of which was motivated by the work of [1] and [2].  We
term the luminosity estimator ``Cepheid-like'' in that it can be used to infer
luminosities $L$ and luminosity distances for the long-duration bursts from the
variabilities $V$ of their light curves alone.  A preliminary application of
this luminosity estimator to 907 long-duration bursts appears in [3].

We discuss the construction of our measure of $V$ in \S 2.  In \S 3, we discuss
our expansion of the original [1] sample of 7 bursts to include a total of 20
bursts, including 13 BATSE bursts, 5 {\it Wind}/KONUS bursts, 1 {\it
Ulysses}/GRB burst, and 1 NEAR/XGRS burst.  Also in \S 3 we discuss the
construction of our luminosity estimator.

\begin{figure}[t]
\begin{center}
\includegraphics[width=1\textwidth]{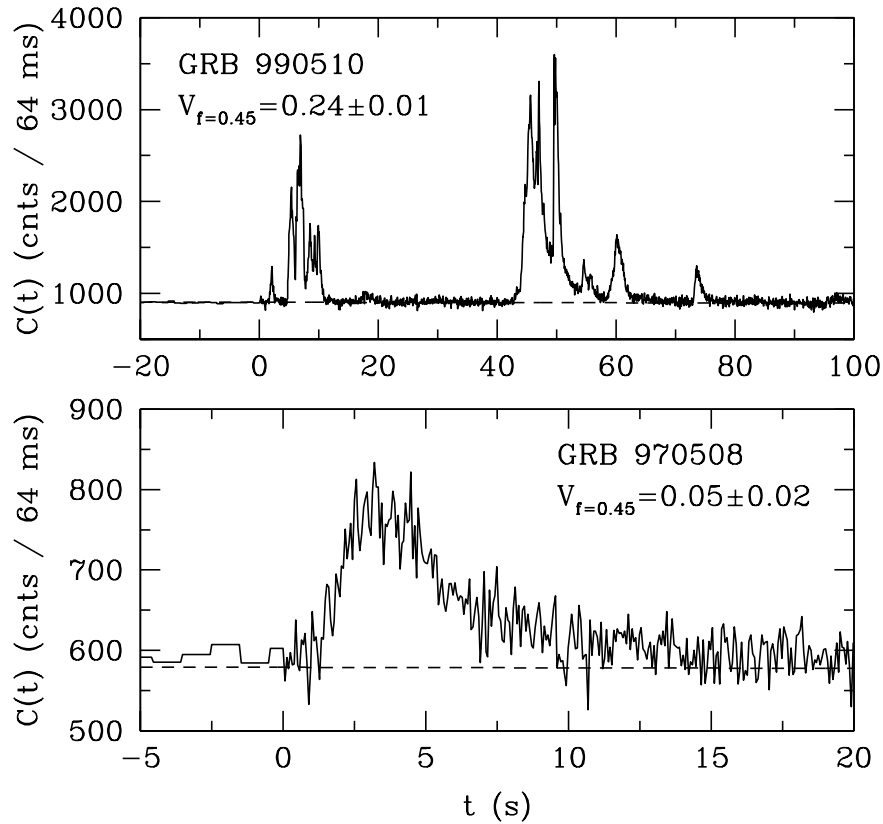}
\end{center}
\caption[]{The $> 25$ keV light curves of the most (GRB 990510)
and least (GRB 970508) variable cosmological BATSE bursts in our sample.  In the
case of GRB 990510 ($z = 1.619$), we find that $V = 0.24 \pm 0.01$.  In
the case of GRB 970508 ($z = 0.835$), we find that $V = 0.05 \pm
0.02$.}
\label{figure1}
\end{figure}

\section{The Variability Measure}

Qualitatively, $V$ is computed by taking the difference of the light curve and a
smoothed version of the light curve, squaring this difference, summing the
squared difference over time intervals, and appropriately normalizing the
result.  We rigorously construct $V$ in [4].  We require it to have the
following properties:  (1) we define it in terms of physical, source-frame
quantities, as opposed to measured, observer-frame quantities; (2) when
converted to observer-frame quantities, all strong dependences on redshift and
other difficult or impossible to measure quantities cancel out; (3) it is not
biased by instrumental binning of the light curve, despite cosmological time
dilation and the narrowing of the light curve's temporal substructure at higher
energies [5]; (4) it is not biased by Poisson noise, and consequently can be
applied to faint bursts; and (5) it is robust; i.e., similar light curves always
yield similar variabilities.  Also in [4], we derive an expression for the
statistical uncertainty in a light curve's measured variability, and we describe
how we combine variability measurements of light curves acquired in different
energy bands into a single measurement of a burst's variability.  We plot the $>
25$ keV light curves of the most and least variable cosmological BATSE bursts in
our sample in Figure 1.

\begin{figure}[t]
\begin{center}
\includegraphics[width=1\textwidth]{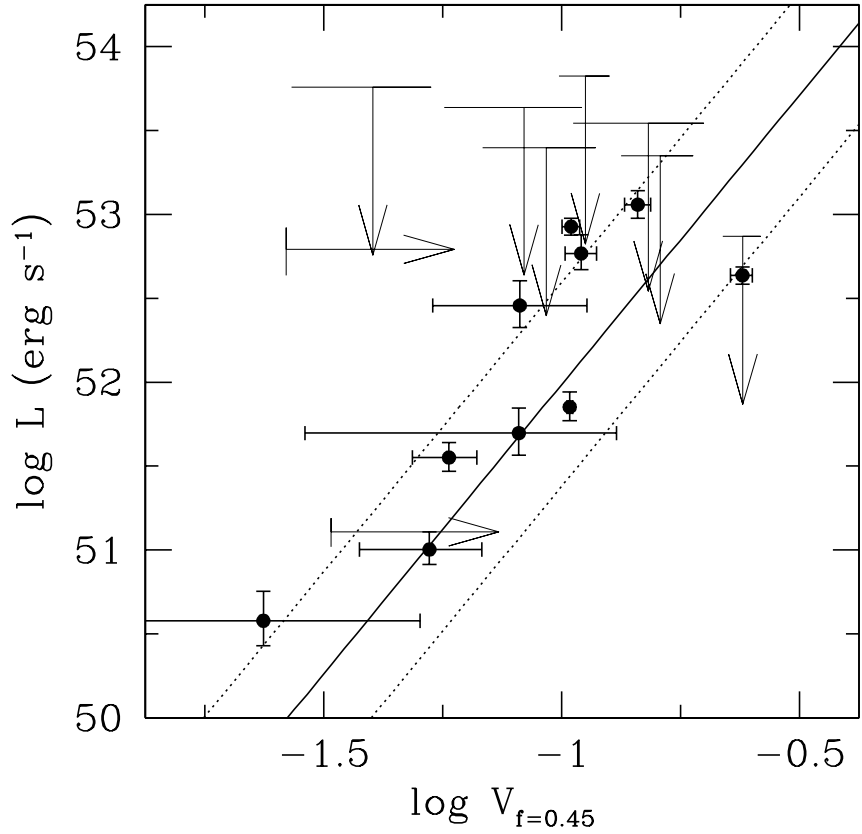}
\end{center}
\caption[]{The variabilities $V$ and
isotropic-equivalent peak photon luminosities $L$ between source-frame energies 
100 and 1000 keV (see [4]) of the bursts in our sample,
excluding GRB 980425.  The solid and dotted lines mark the center and 1 $\sigma$
widths of the best-fit model distribution of these bursts in the
$\log{L}$-$\log{V}$ plane.}
\label{figure2}
\end{figure}

\section{The Luminosity Estimator}

We list our sample of 20 bursts in Table 1 of [4]; it consists of every burst
for which redshift information is currently available.  Spectroscopic redshifts,
peak fluxes, and high resolution light curves are available for 11 of these
bursts; partial information is available for the remaining 9 bursts. We 
rigorously construct the luminosity estimator in [4], applying the Bayesian 
inference formalism developed by [6].  We plot the data and best-fit model of 
the distribution of these data in the $\log{L}$-$\log{V}$ plane in Figure 2.

%

\end{document}